\documentclass[a4paper]{article}
\usepackage{INTERSPEECH2021}
\usepackage{amsmath,graphicx}
\usepackage{array}
\usepackage{lipsum}
\usepackage{graphicx}
\usepackage{epstopdf}
\usepackage{url}
\usepackage{bm}
\usepackage{hhline}
\usepackage{caption}
\usepackage{subcaption}
\usepackage{todonotes}
\usepackage{hyperref}
\usepackage{multirow}

\newcolumntype{x}[1]{>{\centering\let\newline\\\arraybackslash\hspace{0pt}}p{#1}}

\title{Leveraging speaker attribute information using multi task learning for speaker verification and diarization}
\name{Chau Luu, Peter Bell, Steve Renals\thanks{Supported by an EPSRC iCASE studentship collaboration with the BBC.}}
\address{Centre for Speech Technology Research,  University of Edinburgh, UK}
\email{\{chau.luu, peter.bell, s.renals\}@ed.ac.uk}

\begin{document}
%
\maketitle
\begin{abstract}
Deep speaker embeddings have become the leading method for encoding speaker identity in speaker recognition tasks. The embedding space should ideally capture the variations between all possible speakers, encoding the multiple acoustic aspects that make up a speaker's identity, whilst being robust to non-speaker acoustic variation.  Deep speaker embeddings are normally trained discriminatively, predicting speaker identity labels on the training data.  We hypothesise that additionally predicting speaker-related auxiliary variables -- such as age and nationality -- may yield representations that are better able to generalise to unseen speakers.  We propose a framework for making use of auxiliary label information, even when it is only available for speech corpora mismatched to the target application.  On a test set of US Supreme Court recordings, we show that by leveraging two additional forms of speaker attribute information derived respectively from the matched training data, and VoxCeleb corpus, we improve the performance of our deep speaker embeddings for both verification and diarization tasks, achieving a relative improvement of 26.2\% in DER and 6.7\% in EER compared to baselines using speaker labels only.  This improvement is obtained despite the auxiliary labels having been scraped from the web and being potentially noisy.
\end{abstract}


\section{Introduction}
\label{sec:intro}

Obtaining speaker discriminative features is an important step for many speaker recognition tasks, such as speaker verification and speaker diarization. In recent years, extracting speaker embeddings from the intermediate layer of a neural network has become the state-of-the-art method for both tasks~\cite{Snyder2018,Sell2018}, outperforming the historically successful i-vector method~\cite{Dehak2011}.

The ideal speaker embedding space should discriminatively capture the variations between speakers in manner that generalises well to unseen speakers.  The properties of speech encapsulating a speaker's identity are many and varied, including factors related to the underlying physical properties of the vocal apparatus (including factors related to gender, age and some medical conditions), as well as properties related to accent, dialect, native language and sociolect (education level, which affects things like lexicon, syntax and stylistics).  When humans hear speech from a unfamiliar speaker, we intuitively infer many of these factors in the process of forming a mental picture of the speaker.  This technique is used in the field of forensic phonetics and acoustics~\cite{Jessen2007,Hansen2015}, where speaker classification is performed by human experts for suspects in criminal cases.  It is reasonable to expect therefore that a good speaker embedding should encode the multiple contributing factors that constitute our notion of a speaker identity.  
Indeed, previous work has shown that both deep speaker embeddings and i-vectors encode a wide variety of information and meta-information about speakers and utterances, such as speaker emotion~\cite{Williams2019,Pappagari2020}, accent and language~\cite{Maiti2020} or speaker gender, channel and transcription information \cite{Raj2019}.

This work hypothesises that explicitly encouraging deep speaker embeddings to encode multiple speaker attributes may result in a more descriptive embeddings space that is better able capture speaker variation, particularly between unseen speakers.  This is particularly important for the task of diarization where typically all speakers are previously unseen.  For this goal we propose to use multi-task learning (MTL)~\cite{Caruana1998} to leverage speaker attribute information when training deep speaker embeddings.  A challenge of the use of attribute labels is that data with the relevant labels may be hard to find, particularly from the domain of interest.  We solve this problem by designing a system that is able to make use of attribute labels from out-of-domain data.  Furthermore, because the attribute labels are used only as a secondary task in MTL, and discarded at test time, our method is robust to noise in the labels, which allows us to obtain them by means of web scraping.  We demonstrate the method by web scraping nationality labels for VoxCeleb 2 \cite{Nagraniy2017,Chung2018} and age information for speakers in recordings of the US Supreme Court, showing that training on nationality and age classification tasks in conjunction with the standard speaker classification can improve deep speaker embedding performance on both verification and diarization tasks. Experimental and web scraping code has been made available\footnote{\url{https://github.com/cvqluu/MTL-Speaker-Embeddings}}.

The concept of MTL revolves around the concept that machine learning models that may be used to solve different problems using the same data can benefit from sharing a common representation.  In the field of automatic speech recognition (ASR), the work of \cite{Parveen2003} found improvement in increasing the robustness of a hybrid RNN/HMM system by performing speech enhancement as an additional task to classification, with both tasks relying on the same hidden representation, while the work of~\cite{bell2017multitask} found that training to simultaneously predict both context-dependent and context-independent targets regularly improved performance in an ASR setting.

Previous works implementing MTL for speaker recognition specifically have explored using the word spoken in the utterance as an additional task in training deep speaker embeddings to increase verification performance~\cite{Dey2018}. Likewise,~\cite{Liu2019} found that learning to additionally classify the phonetic information also improved performance. While these previous works managed to improve speaker embedding robustness by utilizing transcription information to inform the learned speech representation, there is a risk that these approaches result in embeddings that are more sensitive to phonetic variation, a nuisance factor when performing speaker diarization; this work instead proposes to utilize information that is explicitly connected to the speaker identity to establish a more descriptive and robust speaker embedding space. 




\section{Multi-task learning}
\label{sec:mtl}

\begin{figure}[!tb]
    \centering
    \captionsetup{width=\linewidth}
    \includegraphics[width=\columnwidth]{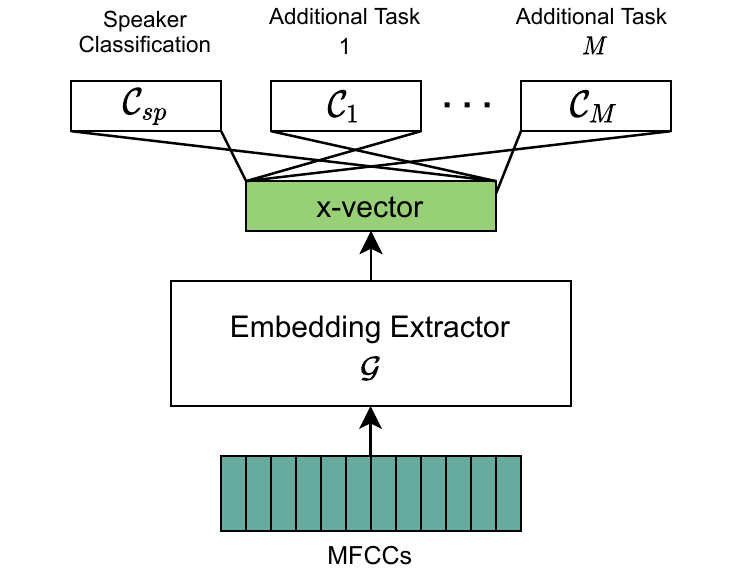}
    \caption{\label{fig:mtl_diag}{An example of a deep speaker embedding extractor trained with multiple tasks.}}
    \vspace{-5mm}
\end{figure}

The concept of multi-task learning is that machine learning models may benefit from sharing the same representations when solving different tasks on the same data. In the context of deep learning, this typically means initial layers of a neural network are shared between tasks, after which task specific layers act upon these shared representation of the input data.

We apply MTL in training the speaker embedding extractor, such as the x-vector network \cite{Snyder2018}, which takes acoustic features as input and performs speaker classification. The speaker embeddings are extracted from an intermediate layer in this network. If the layers up until the extracted embedding are considered to be the embedding extractor, one can consider the remaining layers to be a task specific `head'. The standard x-vector network has a single task specific head, a feed forward network performing speaker classification. Multi task learning can be applied by adding additional, separate task-specific heads with their own loss functions which also act on the embedding.



All components of the system, meaning the embedding extractor $\mathcal{G}$, the speaker classification head $\mathcal{C}_{\text{speaker}}$, and each additional task head $\mathcal{C}_{\text{m}}$ for some $M$ number of additional tasks, are trained as a whole. This is illustrated in Figure \ref{fig:mtl_diag}. Each task head produces a loss based on the specific task. For example, the speaker classification loss is likely to be the standard cross entropy loss or angular penalty loss \cite{Wang2018}. An example for an additional task could be to predict the speaker's age. This could be a regression task, predicting the age and utilizing a mean squared error loss, or a classification task, predicting the age category out of some discretized bins.

Starting from the assumption that a speaker classification loss will always be applied, for $M$ additional tasks, their losses $\mathcal{L}_m$ can be combined in the following manner,

\begin{equation}
    \mathcal{L_{\text{multi-task}}} = \mathcal{L}_{\text{speaker}} + \sum^M_{i=1} \lambda_{m}  \mathcal{L}_{\text{m}}
\end{equation}

\noindent where each additional task loss $\mathcal{L}_{\text{m}}$ is weighted by some chosen loss weighting $\lambda_{m}$, relative to the speaker loss $\mathcal{L}_{\text{speaker}}$. With this formulation, it is possible to consider many additional tasks based on a variety meta-information, but in this work we typically consider only a single additional task. 


Having the embedding space explicitly predict properties which we expect to be informative of identity should lead to robustness in that aspect.  For example, if accent is used as a task, we encourage the space to structure speaker identity in a fashion we are confident will lead to an appropriate embedding for a new speaker with an accent seen during training.  Training only on speaker label may lead to this structure appearing naturally, but previous work has shown that this training objective alone may lead to the embedding space capturing non-speaker related information, such as channel information \cite{Luu2020, huh_augmentation_2020} that we would prefer to model to be invariant to. 



\section{Speaker Attributes}

In order to explore leveraging speaker attributes for MTL, we require the availability of data which has this meta-information.  Our insight here is that we can opportunistically take advantage of such information when it is happens to be available, by applying MTL in a domain where additional attribute labels are available, followed by transfer learning to adapt the resulting embedding to be more suitable for the intended target domain. 

We illustrate the method using two datasets, each with an additional attribute label: the Supreme Court of the United States (SCOTUS) corpus, for which we obtained information on speaker age; and VoxCeleb, which has obtainable information about speaker nationality.  We find SCOTUS to be more challenging than VoxCeleb for diarization and hence adopt it as our primary task, although speaker verification results on VoxCeleb are also presented.

It should be noted that both sets of attribute labels used in this work were obtained via web scraping, and thus are more susceptible to noise and inaccuracy than labels obtained via other means.  Nevertheless, we find that they can provide useful information about how speaker identity should be structured in the embedding space.  In general, while noisy labels generally do degrade performance in most tasks, it is still possible to train under this paradigm and yield positive results \cite{xie_disturblabel_2016,song_learning_2020}. 


\subsection{Age: SCOTUS Corpus}
\label{sec:scotus}

The Supreme Court of the United States is the highest level court in the United States. Audio recordings and transcriptions of SCOTUS oral arguments have been made available via the Oyez project\footnote{\url{https://www.oyez.org/about}}. This corpus is unusual in that it features certain speakers across many years. These speakers are the justices (judges) of the Supreme Court, who serve indefinitely on the court until their retirement. The average length of a supreme court justice's tenure at time of writing is approximately 17 years. The phenomenon of how a human voice can change due to the effects of age is fairly well known \cite{Mueller1997}, with degradation in the vocal folds sometimes leading to changes in the fundamental frequency of the voice, and other qualities in vocal delivery, such as jitter, shimmer and volume. Severe cases can lead to diagnoses of vocal disorders \cite{Rapoport2018}. These effects on a speaker's voice makes age a fascinating aspect to explore of what contributes to speaker identity.

\begin{figure}[!tb]
    \centering
    \captionsetup{width=\linewidth}
    \includegraphics[width=\columnwidth]{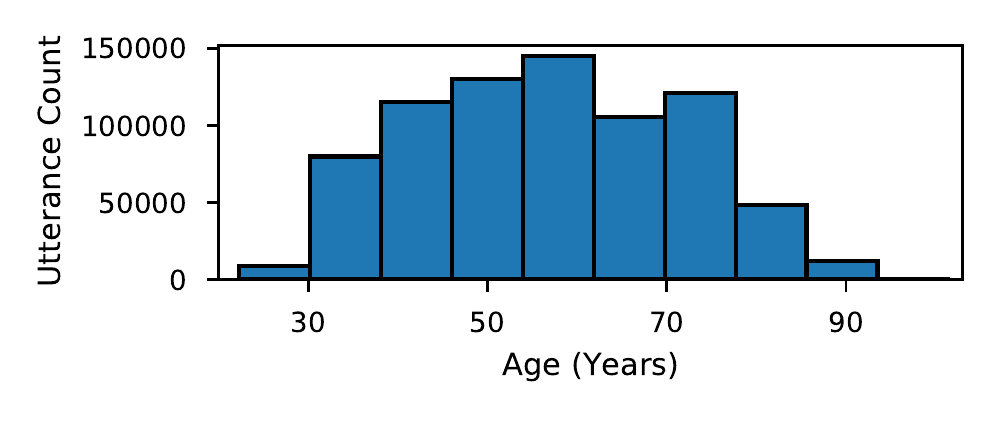}
    \vspace{-5mm}
    \caption{\label{fig:scotus_age_dist}{The age distribution of utterances in the SCOTUS corpus, split into 10 bins.}}
    \vspace{-5mm}
\end{figure}

For this work, 1022 digital recordings with 913 unique speakers across 1032 hours of audio were considered. For most speakers, approximate ages were obtained for speakers based on their date of qualification to practice law, which was obtained via web scraping. The interested reader can refer to the code repository for more information. One can see the distribution of ages of utterances in Figure \ref{fig:scotus_age_dist}, with an average age of around 56.

\subsection{Nationality: VoxCeleb}

VoxCeleb 1 and 2 \cite{Nagraniy2017,Chung2018} are speaker recognition datasets featuring celebrity speakers, meaning certain speaker attributes can be found in the public domain. Speaker nationality labels can be web scraped from sources such as Wikipedia and are a proxy for speaker accent, an attribute which is clearly informative of speaker identity.

\begin{figure}[tb]
    \centering
    \captionsetup{width=\linewidth}
    \includegraphics[width=0.7\columnwidth]{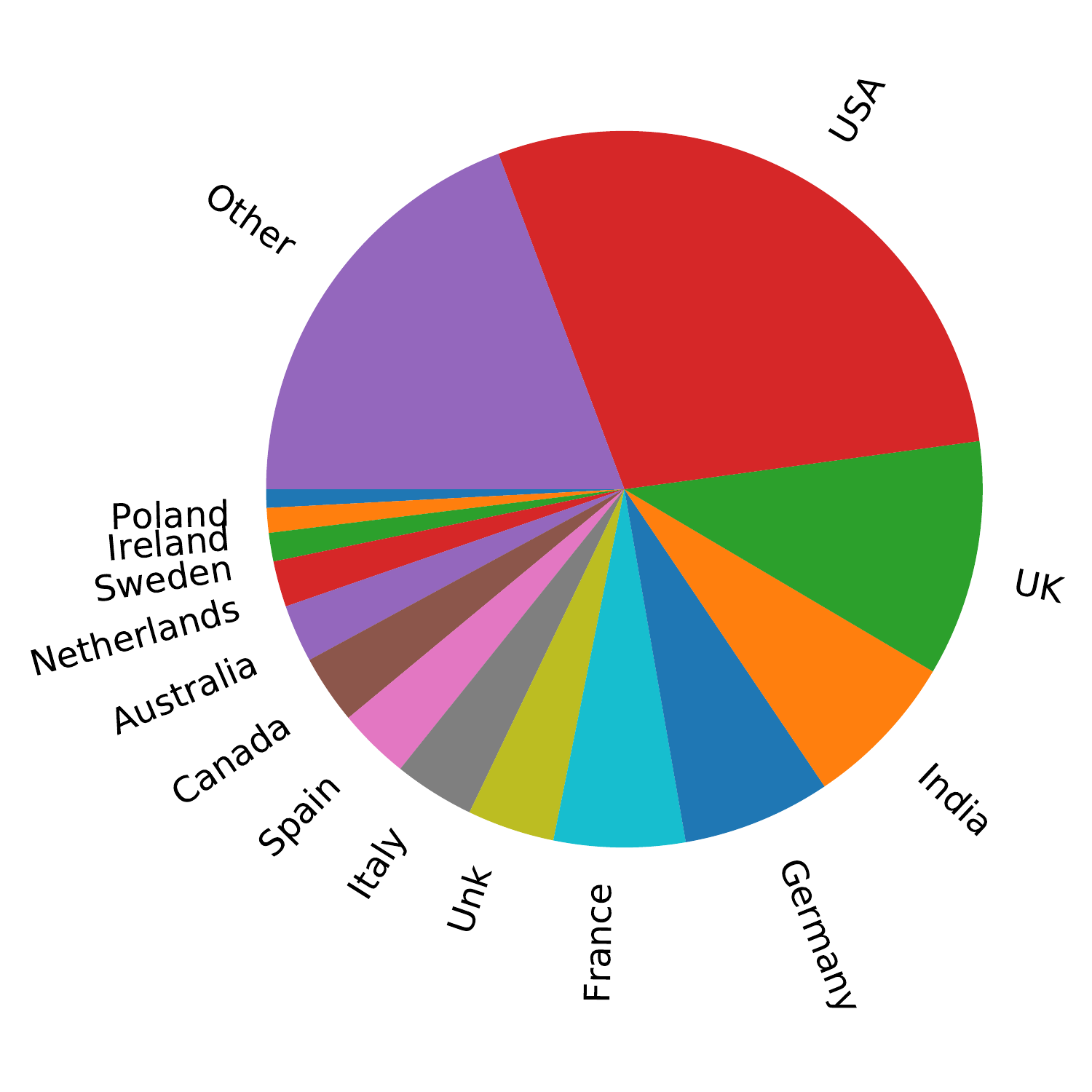}
    \vspace{-5mm}
    \caption{\label{fig:vox_dist}{The nationality distribution of the 5994 speakers in VoxCeleb 2's training set. Only nationalities with more than 50 are shown individually in this figure, with the rest being grouped as `Other'.}}
    \vspace{-5mm}
\end{figure}



\section{Experimental Setup}
\label{sec:exp}

Our primary verification and diarization experiments were performed on SCOTUS corpus, whilst embeddings were also trained on VoxCeleb. 
For the training and evaluation on SCOTUS, utterances were split into train and test sets by recording, at an approximate $80\%$ train proportion, and ensuring that the train and test distributions for age using 10 uniformly spaced bins were approximately similar. In order to evaluate the SCOTUS data as a verification task, 15 positive and negative trials were selected for each speaker in the test recordings, excluding any speakers seen in the training set from these trials. Cosine similarity was used to score all embeddings.

The speaker embedding extractors for both VoxCeleb and SCOTUS followed the original x-vector architecture, up until the embedding layer which had 256 hidden units. For classification heads that utilized the standard cross entropy loss, these had a similar architecture to the x-vector network, having two hidden layers also of dimension 256 before being projected to the number of classes. The non-linearity used throughout was Leaky ReLU. For classification heads that used an angular penalty loss, these were simply a single affine matrix on top of the embedding layer that projected into the number of classes, using the CosFace~\cite{Wang2018} loss.
Embedding extractors were trained with different configurations of classification heads, and verification and diarization performance was evaluated.  


We also performed contrastive experiments with randomly shuffling the labels of the additional tasks -- such as age -- to eliminate the possibility any kind of positive regularization effect that the additional task may have irrespective of the information in the labels.

For all embedding extractor training setups for SCOTUS, regardless of the number of classification heads, networks were trained for 50,000 iterations on 350 frames of 30-dimensional MFCCs, with batch size 500, using a small held out set of training utterances for validation. Stochastic gradient descent was used with learning rate 0.2 and momentum 0.5.

For diarization of the SCOTUS corpus, embeddings were extracted for every 1.5s with 0.75s overlap using reference speaker activity detection segmentation, using cosine similarity for scoring and using agglomerative hierarchical clustering to the oracle number of speakers. Due to the supreme court justices appearing in both train and test recordings, the  diarization error rate was evaluated for two scenarios: The first evaluation scenario was standard in that all the speech segments were scored, including the speakers which appeared in the training set. The second evaluation scenario was to only score portions of the speech in which unseen speakers were talking.

Speaker embedding extractors for VoxCeleb were trained on the VoxCeleb 2 training set, with 5994 speakers, augmented in the standard Kaldi~\cite{Povey_ASRU2011} fashion with babble, music and background noises along with reverberation as in \cite{Snyder2018}. Speakers who were the only members of their nationality in the training set were grouped into the same class as the speakers who could not have their nationality scraped, yielding 102 nationality classes. Note that not all of these are shown in Figure \ref{fig:vox_dist}. Networks were trained for 100,000 iterations with the same batch size and optimization settings as the SCOTUS models.

When evaluating models trained on VoxCeleb on SCOTUS, the VoxCeleb models was were fine-tuned for 5000 iterations on the SCOTUS training set, varying whether or not the full network or only the last linear layer of the extractor was fine-tuned, along with whether or not age in addition to speaker labels were trained on during the fine-tuning. For the first 1000 iterations of fine-tuning, all embedding extractor parameters were frozen to allow the freshly initialized classification head(s) to fit to the new data.

\section{Results and discussions}
\label{sec:results}

\begin{table*}[t]
  \centering
  \begin{tabular}{|c| m{4cm} || c | c c c || c |}
  \hline
   \multirow{3}{*}{\shortstack{Training\\Set}} & \multirow{3}{*}{Model} & \multirow{3}{*}{\shortstack{SCOTUS\\Fine-tune label set}} & \multicolumn{3}{c||}{SCOTUS} & VoxCeleb \\
  \cline{4-7}
    & & & \multirow{2}{*}{EER} & \multicolumn{2}{c||}{DER} & \multirow{2}{*}{EER} \\ 
   &  & &    &  All & Unseen &  \\
  \hline
  \multirow{8}{*}{\rotatebox[origin=c]{90}{SCOTUS}} & Only Speaker & - & 3.14\% & 27.58\% & 19.75\% & - \\
  & Speaker + Random labels &- & 3.78\% &  27.87\% & 19.14\% & - \\
  & Speaker + Age & - & 2.68\% &  26.14\% & 18.02\% & - \\

  \cline{2-7}
  & Only Speaker (CosFace) & - & 2.71\% &  26.51\% & 19.75\% & - \\

  & Speaker (CosFace) + Age & - & \textbf{2.62\%} & \textbf{21.80\%} & \textbf{14.08\%} &- \\

 \cline{2-7}
  & Only Age & - & 3.99\% &  37.08\% & 26.18\% & - \\
  \cline{2-7}
  & Only Gender & - & 19.42\% &  58.02\% & 44.78\% & - \\
  \cline{2-7}
  & Only Random & - & 23.31\% &  69.13\% & 48.97\% & - \\
  \hhline{|=|=#=|===#=|}
 
  \multirow{10}{*}{\rotatebox[origin=c]{90}{VoxCeleb 2}} & Only Speaker (Baseline) & (Last Linear) Sp. & 2.26\% &  20.09\% & 14.07\% & 3.04\% \\
  & Speaker + Random & (Last Linear) Sp. & 2.98\% &  22.18\% & 16.32\% & 4.32\% \\
  & Speaker + Nationality & (Last Linear) Sp. & 1.99\% &  18.74 	\% & 13.54\% & \textbf{2.95\%} \\
  \cline{2-7}
     &  Only Speaker & (LL) Sp. + Age & 2.00\% & 17.81\% & 12.44\%  & 3.04\% \\
  & Speaker + Nationality  & (LL) Sp. + Age & 1.89\% & \textbf{14.82\%} & \textbf{10.45\%}  &\textbf{2.95\%} \\
  \cline{2-7}
&  Only Speaker (Baseline) & (Full) Sp.  & 1.63\% & 29.56\% & 20.02\%  & 3.04\% \\
  & Speaker + Nationality  & (Full) Sp. & 1.57\% & 25.04\% & 16.93\%  & \textbf{2.95\%} \\
  \cline{2-7}
    &  Only Speaker & (Full) Sp. + Age & 1.57\% & 25.98\% & 17.44\%  & 3.04\% \\
  & Speaker + Nationality  & (Full) Sp. + Age  & \textbf{1.52\%} & 19.77\% & 13.57\% & \textbf{2.95\%} \\
  \cline{2-7}
  & Only Nationality & - & 17.68\% & 57.97\% & 42.78\% & 13.38\%\\
  \hline 
  \end{tabular}
  \caption{\label{tab:scotus_multi} Verification and diarization performance on SCOTUS and VoxCeleb for various models with multitask learning objectives.}
\end{table*}

All results are shown in Table \ref{tab:scotus_multi}.  
Results for adding a 10-class age classification task to speaker embedding extractors for the SCOTUS corpus can be seen in the first portion of the table, separated into the models trained with cross entropy loss on the speaker classification head, and the models trained with CosFace loss on the speaker classification head. The Diarization Error Rate (DER) results are split into two scoring scenarios, `All' indicates that all speech was scored and `Unseen' indicates that only speech segments from unseen speakers was scored. Adding gender classification was not found to have any positive effect, and thus these results have been omitted for brevity.

For the standard cross entropy loss, for configurations of the age loss with $\lambda_{\text{age}}$=0.5, the verification and diarization performance was improved over the baseline. (For brevity, not all values of $\lambda$ are shown in the table).  By contrast, both the control experiment of randomly shuffled labels and the control experiment featuring completely random speaker labels yielded no improvement.

The SCOTUS trained models also feature an experiment in which the only classification head was the Age classification head, and this performs surprisingly well on the speaker verification and diarization tasks, despite only being trained to distinguish between 10 age categories. This suggests that in order to predict age, some knowledge of speaker identity is also required. Although it is not shown on the table, the age accuracy of the non-CosFace networks trained with both speaker and age outperformed that of training on age alone, suggesting that performing tasks in combination was able to improve both tasks. This is supported by previous MTL literature, which indicates training on multiple tasks may be helpful to each individually. 

The addition of gender classification not improving results is unsurprising, considering male and female voices can largely be distinguished based on their fundamental frequency (F0) \cite{Jessen2007, Pernet2012}, and thus this does not add much discriminatory power to the primary goal of speaker recognition.

For SCOTUS trained models, it is clear from Table \ref{tab:scotus_multi} that CosFace improves the speaker verification and diarization performance over the standard cross entropy loss, with $\lambda_{\text{age}}$ being changed to $0.01$ to account for the change in the relative scale of $\mathcal{L}_{\text{speaker}}$. The addition of the age classification head similarly improves over only using speaker labels, producing the best results for verification and diarization for all the configurations shown, making a relative improvement of 17.8\% in DER and 3.3\% in EER over the `Only Speaker' CosFace baseline.


The performance of VoxCeleb trained models can also be seen in Table \ref{tab:scotus_multi}. Note all speaker classification heads for VoxCeleb models utilized the CosFace loss. The verification performance of these models on VoxCeleb can be seen in the rightmost column, with the addition of the nationality task ($\lambda_{\text{nat}}$=0.05) yielding a 3\% relative performance improvement.

When evaluated on SCOTUS, the fine-tuning of the VoxCeleb models was performed on either the Last Linear (LL) layer of the embedding network, or the whole (Full) network. Either Speaker (Sp.) labels alone were used, or Age was added as an auxiliary fine-tuning task, with $\lambda_{\text{age}}$=0.05. Models which only use speaker labels at all stages of training (marked as `Baseline' in the Table) are improved upon in both verification and diarization by utilizing either Nationality or Age tasks during the primary training stage or the fine-tuning stage respectively. Indeed, the best performance for verification and diarization is found when auxiliary tasks are employed at both stages.

For the best baseline verification model with no additional tasks, a 6.7\% relative improvement in EER is found by using nationality and age (1.63\% $\rightarrow$ 1.52\%), and similarly for diarization, using auxiliary tasks at both stages yields a 26.2\% relative improvement in DER scoring all regions (20.09\% $\rightarrow$ 14.82\%). 

While using both additional tasks yields the best performance, improvements are still found when using a single auxiliary task at either stage of training, suggesting that this technique is still valuable for scenarios in which speaker attribute information is limited or missing from the desired domain.

Overall, these experiments demonstrated that training on auxiliary speaker attribute tasks in addition to speaker classification can yield more robust representations for verification and diarization.

\section{Conclusions}
\label{sec:conclusion}

In this work we showed that training on additional tasks relating to speaker attributes, specifically approximate age and nationality, alongside the standard speaker classification can improve the performance of deep speaker embeddings for both verification and diarization. Future work is planned to investigate training on multiple corpora and multiple tasks simultaneously to extend the improvement in robustness from multi-task training shown here.


\pagebreak

\bibliographystyle{IEEEtran}
\bibliography{strings,refs,references}

\end{document}